\documentclass[12pt]{article}
\usepackage{psfig}
\usepackage{epsf}
\def\ben{\begin{enumerate}}
\def\een{\end{enumerate}}
\def\bit{\begin{itemize}}
\def\eit{\end{itemize}}
\def\ba{\begin{array}}
\def\ea{\end{array}}
\def\bea{\begin{eqnarray}}
\def\eea{\end{eqnarray}}
\def\bq{\begin{quote}}
\def\eq{\end{quote}}

\def\bc {\begin{center}}
\def\ec {\end{center}}
\def\be{\begin{equation}}
\def\ee{\end{equation}}

\def\eg{ {\em e.g.,\ }}

\def\etal{ {\em et al.\ }}

\def\bib{\bibitem}

\def\r {\right}
\def\l {\left}

\def \HU{\left |{H_{u}}\right |}
\def \MHU{m_{H_u}^2}

\def \MHF{m_{1/2}}

\def \MCH{m_{\tilde \chi^\pm}}
\def \MGL{m_{\tilde g}}

\def \msll{{m_{\tilde l}}_L}
\def \mslr{{m_{\tilde l}}_R}
\def\lapp{\mathrel{\rlap{\raise.5ex\hbox{$<$}}
                    {\lower.5ex\hbox{$\sim$}}}}
\def\gapp{\mathrel{\rlap{\raise.5ex\hbox{$>$}}
                    {\lower.5ex\hbox{$\sim$}}}}

\def \g2sq{g_2^2}

\def \QA{\hat Q_A(H_u)}

\def\issue(#1,#2,#3){#1 (#3) #2} 

\def\opcit(#1){ {\em op. cit.}, #1}

\def\APP(#1,#2,#3){Acta Phys.\ Polon.\ \issue(#1,#2,#3)}
\def\ARNPS(#1,#2,#3){Ann.\ Rev.\ Nucl.\ Part.\ Sci.\ \issue(#1,#2,#3)}
\def\CPC(#1,#2,#3){Comp.\ Phys.\ Comm.\ \issue(#1,#2,#3)}
\def\CIP(#1,#2,#3){Comput.\ Phys.\ \issue(#1,#2,#3)}
\def\EPJC(#1,#2,#3){Eur.\ Phys.\ J.\ C\ \issue(#1,#2,#3)}
\def\EPJD(#1,#2,#3){Eur.\ Phys.\ J. Direct\ C\ \issue(#1,#2,#3)}
\def\IEEETNS(#1,#2,#3){IEEE Trans.\ Nucl.\ Sci.\ \issue(#1,#2,#3)}
\def\IJMP(#1,#2,#3){Int.\ J.\ Mod.\ Phys. \issue(#1,#2,#3)}
\def\MPL(#1,#2,#3){Mod.\ Phys.\ Lett.\ \issue(#1,#2,#3)}
\def\NP(#1,#2,#3){Nucl.\ Phys.\ \issue(#1,#2,#3)}
\def\NIM(#1,#2,#3){Nucl.\ Instrum.\ Meth.\ \issue(#1,#2,#3)}
\def\PL(#1,#2,#3){Phys.\ Lett.\ \issue(#1,#2,#3)}
\def\PRD(#1,#2,#3){Phys.\ Rev.\ D \issue(#1,#2,#3)}
\def\PRL(#1,#2,#3){Phys.\ Rev.\ Lett.\ \issue(#1,#2,#3)}
\def\SJNP(#1,#2,#3){Sov.\ J. Nucl.\ Phys.\ \issue(#1,#2,#3)}
\def\ZPC(#1,#2,#3){Zeit.\ Phys.\ C \issue(#1,#2,#3)}
\def\JHEP(#1,#2,#3){JHEP\ \issue(#1,#2,#3)}

\begin{document}
\begin{center}
{\Large{\bf LEP Data and the Stability of the Potential 
Confront  the mSUGRA Model}}
\vskip 25pt
{\large Amitava Datta $^{a,}$
\footnote{Electronic address: adatta@juphys.ernet.in}}
and 
{\large Abhijit Samanta $^{b,c,}$
\footnote{Electronic address: abhijit@anp.saha.ernet.in}}
\vskip 10pt
$^a${\it Department of Physics, Jadavpur University, Kolkata - 700 032, India}\\
$^b$ {\it Saha Institute of Nuclear Physics, 1/AF Bidhan Nagar, Kolkata
700064, India  \\
$^c$ Department of Physics, University of Calcutta, Kolkata - 700 009, India}
\end{center}
\vskip 20pt

\begin{abstract}

The requirement that the supersymmetric scalar potential be stable
in the minimal supergravity (mSUGRA) model
 imposes an upper bound on the universal gaugino mass 
$\MHF$ as function of the common scalar mass $m_0$ . Combining this with
the experimental lower bound on $\MHF$ from LEP data,
we find  a new lower bound on 
$m_0$, stronger than the one that comes from experimental 
data alone. If the  corresponding upper limits on the 
superparticle masses,
derived  in this letter, are found to be violated at Tevatron Run II
or at the LHC, it would imply that we are living on a false vacuum.
Special attention has been paid in estimating the uncertainties in these
predictions due to the choice of the renormalization scale.
The implications of our limits for the constraints obtained by indirect
methods( SUSY dark matter, g - 2 of the muon, $ b 
\rightarrow s \gamma$.... ) are briefly  discussed. 
\end{abstract}
PACS no: 12.60.Jv, 14.80.Ly, 14.80.Cp
\setcounter{footnote}{0}
\renewcommand{\thefootnote}{\arabic{footnote}}

Currently Supersymmetric theories are among the best motivated extensions
of the Standard Model (SM)\cite{susy-review}. The most general one 
at low energy is the Minimal Supersymmetric extension of the 
Standard Model (MSSM). However, its phenomenological study is almost 
impossible due to large number of free parameters.
Almost all of them arise due to our failure to discover the precise 
mechanism of supersymmetry (SUSY) breaking. Currently there are several 
popular models of supersymmetry breaking. The theoretical assumptions
in these  models reduce the parameter space. 
The most well-studied model is the minimal
supergravity (mSUGRA) model \cite{sugra} with radiative electroweak 
symmetry breaking \cite{rewsb}. This model has only five free parameters.
They are the common scalar mass ($m_0$), the common gaugino mass ($\MHF$),
the common trilinear coupling ($A_0$), the ratio of vacuum expectation
values of two Higgs field ($\tan\beta$) and the sign of $\mu$, the
higgsino mass parameter.

The mSUGRA model has been confronted with the data from LEP 
as well as from Run I of the Tevatron collider. Such efforts have 
resulted in
some useful lower bounds on these parameters most notably on 
$m_0$ and $\MHF$\cite{aleph,lepbound,tevatronbound,opal}. In this paper
we shall be concerned mainly with the limits obtained by the ALEPH
collaboration on the mSUGRA parameter space \cite{aleph}. The results
obtained by the other LEP groups are similar but some of them do not 
strictly follow the mSUGRA scenario.

 Unfortunately mSUGRA does not predict  quantitative  upper bounds
 on these parameters, which could make this model falsifiable 
in the near future. Of course there are upper bounds 
based on naturalness arguments\cite{naturalness} . These 
bounds, however, crucially depends on the value of the 
 fine tuning parameter which, though intuitively appealing,
 is rather difficult to quantify.

It is, therefore, rather tempting to reexamine the constraints obtained 
from the stability of the supersymmetric scalar 
potential\cite{oldufb,casas,anirban}.
It has been demonstrated in the past that these constraints leads to
upper bounds on $\MHF$ as a function of $m_0$\cite{casas,anirban}. Useful
constraints also emerge within the framework of anomally mediated 
supersymmetry breaking and other models\cite{sourav}. 

Admittedly these bounds can be evaded by assuming that we live 
in a false vacuum with a life time larger than the age of the universe
\cite{falsevacuum}. Yet such bounds are important. 
If they are found to be  violated  after the discovery of SUSY be it at 
Tevatron
Run II or at the LHC, it must be accepted within the mSGRA scenario that the 
universe is indeed built on a false vacuum.

The unbounded from below (UFB)\cite{casas} constraint imposes an upper 
limit on $\MHF$ (denoted by $\MHF^{\rm max}$)
as function of $m_0$ (to be explained below). Combining this with
the experimental lower bound denoted by  
$\MHF^{\rm min}$ from LEP data \cite{aleph},
we find a strong  lower bound on 
$m_0$ (denoted by $m_0^{\rm min}$) stronger than the one that comes from 
experimental data alone. This can be translated into a stronger lower
bound on slepton mass. Using the lower bound (from experimental data)
and upper bound (from the UFB bound) on $\MHF$ for a given $m_0$
we derive upper and lower bounds on various super particle masses for
fixed slepton masses. This prediction can be tested in the next 
round of experiments. 

To make this letter self contained, we briefly discuss the most important, 
model independent unbounded from below - 3 
(UFB3) constraint obtained
by considering a certain direction in the field space\cite{casas}.
The scalar potential in this direction is given by
\be
V_{UFB3}=[\MHU+m_{L_{i}}^{2}]\l|H_u\r|^2+\frac{\left|\mu\right|}
{\lambda_{E_j}}[m_{L_j}^{2}
+m_{E_{j}}^{2}+m_{L_{i}}^{2}]\l|H_u\r|-
\frac{2{m_{L_i}}^4}{{g^\prime}^2+g^2},
\label{ufb3}
\ee
where $g^\prime$ and $g$ are normalised gauge couplings of $U(1)$ and
$SU(2)$ respectively, $H_u$ and $H_d$ are the neutral components of 
the two Higgs doublets,
$L_i$ and $E_j$ are the scalar partners of the 
leptons belonging to the $SU(2)_L$  doublet
and singlet respectively, $\lambda_{E_j}$ is a Yukawa coupling and $i,j$
are generation indices. Here $ i\neq j$. 

Note that we could substitute squarks for sleptons in Eq. \ref{ufb3}, in 
which case $i=j$ is allowed. The constraints on the parameter space 
arise from the requirement 
\be
V_{UFB3}(Q=\hat{Q})>V_0^{\rm min}(Q=M_{S})\label{ufb3condition}
\ee
where $V_0^{\rm min}$ is the electroweak symmetry breaking minimum of
the  scalar potential evaluated at the SUSY breaking scale Q =$M_S$
and the scale  
$\hat{Q}$ is chosen to be\\ $\hat Q\sim Max( g_{2}\left|E_j\right|,  
g_{2}\HU, \lambda_{top}\HU, g_{2}\left|L_{i}\right|,
M_{S}) \sim M(\phi)^{\rm max},$ where $M(\phi)^{\rm max}$ is the largest
eigen value of the field dependent mass matrices. The VEVs
$|E_j|$ and $|L_i|$ are determined by $H_u$ and some model parameters.
 The relevant equations can be found in \cite{casas}. The UFB3 potential in
Eq. \ref{ufb3} is derived from the tree level scalar potential. 
It is well
known that loop corrections to the potential may have important effects
(e.g.,  the predictions from the one loop corrected potential has a
reasonably mild scale dependence) \cite{gamberini}.  
The above choice  of $\hat Q$ is designed
to minimize the magnitude of the loop corrections to the potential 
which are of the form $\sim M^4(\phi)ln\frac{M(\phi)^2}{{\hat Q}^2}$.
At this scale, therefore, the tree level potential by itself gives 
fairly reliable results \cite{casas,gamberini}.
The price that one has to pay for this simplication is a slight
complication in the choice of the scale. While studying $V_{\rm UFB3}$
as a function of $H_u$, one must select the scale at each step since 
it is correlated with $H_u$. 

It should be emphasized that the above prescription only gives an order of
magnitude estimate of the scale $\hat Q$ for a given $H_u$. In order to
carry out practical computations leading to constraints on the parameter 
space, $\hat Q$ is set exactly equal to the
maximum of the quantities given in the parenthesis. In the rest of the paper
this scale will be referred to as the approximate scale $\hat{Q}_A(H_u)$. 
This approximation introduces an 
element of uncertainty in the derived constraints, which  will be discussed
in detail before the numerical results are presented.

As can be seen from Eq. (\ref{ufb3}), the regions of the parameter space, 
where $\MHU$ becomes large negative at the required scale $\QA$,
tends to violate the UFB3 condition (inequality \ref{ufb3condition}). 
This is because the first term of Eq. (\ref{ufb3}) which dominates
for moderate and large values of $|H_u|$,
could be negative in this case. However, the second term in (\ref{ufb3}),
which is positive definite, may become competitive in certain cases 
({\em e.g.}, for $j=1$, when the Yukawa coupling in the denominator is small), 
and a  dangerous minimum may be avoided. 

The UFB3 potential with sleptons (Eq. \ref{ufb3}) was found to yield
the strongest  constraint among all the UFB and 
charge colour breaking (CCB) conditions 
in the low $\tan\beta$ case \cite{casas}. The results were generalized 
for large $\tan\beta$ scenarios in \cite{anirban}.
In order to get the optimum result, one has to take the largest 
$\lambda_{E_{j}}$ in the second term of Eq. (\ref{ufb3}), which leads 
to the choice $E_j = \tilde{\tau}_R$.
Now the restriction $i\neq j$ requires $L_i = \tilde e_L$ or $\tilde \mu_L$
and excludes the choice $\tilde{\tau}_L$. In the low $\tan\beta$ case this 
restriction, however, is of little consequence since all the left sleptons
are degenerate to a very good approximation.

To find the points in the mSUGRA parameter space which violate the
UFB3 constraint (inequality \ref{ufb3condition}), we vary $H_u$ 
from the grand unification scale ($M_G$) 
to $M_Z$. For each $H_u$ and a chosen set of  the five input parameters 
the approximate scale  $\QA$, described 
 after inequality \ref{ufb3condition}, is found by an 
iterative method ( see \cite{anirban} for the details of the procedure). 
Next we evalute the function in Eq. \ref{ufb3}.
At large $H_u$ (i.e., also at large $\QA$ ), the  change of $\MHU$ 
from its boundary value $m_0^2$ at $M_G$ is negligible.
Consequently  the first term
of Eq. \ref{ufb3}, which dominates, is positive and the inequality 
is satisfied. As $H_u$ is decreased $\MHU$ becomes negative.  
Further lowering of $H_u$ may eventually make the first term sufficiently
negative, leading to the violation of inequality \ref{ufb3condition}
for the chosen set of parameters. This set is then excluded.
For some sets of parameters $V_{\rm UFB3}$ is found to be  violated for 
a very narrow range of $\QA$ values. Since $\QA$ is chosen
on the basis of an order of magnitude estimate one may wonder whether 
this choice is the true value of the scale at which 
 conclusions based on the tree level potential are reliable. For some
other sets of parameters, however,  $V_{\rm UFB3}$ remains approximately 
flat for a range of values of $H_u$ ( or $\QA$ ) 
and ineqality \ref{ufb3condition} is  violated for the entire range.
The corresponding parameter set can be ruled out with more confidence.
 For  $H_u$ beyond  this range $V_{\rm UFB3}$ rises above $V_0^{\rm min}$.
It may be inferred that for this range of $\QA$ values higher order effects
are indeed small, the tree level UFB3 potential by itself is fairly scale 
independent  and conclusions based on it  are reliable.  

The variation of $V_{\rm UFB3}$ with $log_{10}(\QA/{\rm GeV})$ is
illustrated in Fig. 1, for $m_0 = 140$GeV,
$ \MHF = 220$ GeV, 
$A_0 = 0,$ $\tan\beta =15$ and $\mu$$>$ 0 by the solid  curve.
The points denoted by the solid traingle and the open circle 
represent $V_0^{\rm min}$ in Eq. (2) for $\MHF$ =  220 and 250 GeV
respectively. For numerical convenience we have plotted the
logarithms of the potential( see the caption of Fig. 1).
The behavour discussed in the 
last paragraph is clearly demonstrated. 
It is to be noted that for this set of parameters
 the UFB3 condition is violated for a very  specific value of $\QA$,
which may not be identical to the true scale where the UFB3 condition
should be tested.  The set of parameters  
under consideration, therefore, cannot be excluded with certainty.

Now we are in a position to illustrate the sensitivity of the excluded
the parameter space to the choice of the scale.
The dashed  curve in Fig. 1 is obtained for  $\MHF=250$ GeV, the
 other parameters being the same as those for the dotted curve.
Now the UFB3 condition is violated for 
a fairly large range of $\QA$. The correspoding parameter space 
can be eliminated with a higher level of confidence.

In view of the above discussions, we have introduced the following 
prescription to estimate the uncertainties in the limits derived in 
this letter due to the choice of 
$\QA$. In the process of obtaining the upper bound on $\MHF$ for a given 
$m_0$, $\MHF$ is increased starting from the LEP lower limit \cite{aleph}. 
At a certain $\MHF$ (= $\MHF^{\rm max}$) 
the UFB3 condition is violated for the first time at a scale $\QA$.
Usually this first violation occurs for a very narrow range of $\QA$ values.
We shall refer to  this  limit  as the optimistic limit (OL). Any 
$\MHF$ above this value violates the UFB3 condition for a larger
range of $\QA$ values. This is so because for a given set of $m_0, A_0, 
\tan\beta$ and sign of $\mu$, a larger $\MHF$ drives $\MHU$ in Eq.
\ref{ufb3} to larger negative values\cite{anirban}. Thus, \eg 
the UFB3 constraint for a larger $\MHF$ will also be violated
at a relatively higher $\QA$.  The  conservative limit (CL) refers to that
$\MHF$ for which
 the UFB3 condition is violated for a range of $\QA$ values, 
$\QA^{\rm min}(=\QA^{\rm max}/10) < \QA \le \QA^{\max}$.
This choice of the range is guided by the conventional understanding
of the order of magnitude. 


In Fig. 2 we present the allowed region of the $m_0 -\MHF$ plane for 
$\tan\beta =15$, $A_0=0$ and $\mu > 0$. In this figure as well as in
the following ones, we have chosen the mSUGRA parameters as in ref 
\cite{aleph}, so that we can combine our theoretical constraints with 
the experimental limits without any ambiguity.
Region I is excluded by the LEP  
lower limit on the light Higgs mass ($m_h$),
which leads to the strongest experimental constraint in the $m_0 - \MHF$ plane
in the mSUGRA model. 
We have used the model independent limits on $m_h$
 as a function of the mixing parameters
obtained by the ALEPH collaboration (see Fig 4 of ref \cite{aleph1})
to reconstruct 
the upper edge of region I, which represents the lower limits
on $\MHF$ for different $m_0$ values. Region IV
is excluded by the requirement that the lightest neutralino be the LSP
which leads to an upper bound on $\MHF$ for a given $m_0$. Thus region II 
and III represent the allowed parameter space (APS) after imposing the 
experimental and the neutralino LSP constraints.

The APS in Fig. 2
is in good agreement with the results of ALEPH collaboration
\cite{aleph}, as can be seen by comparing region I of Fig. 2 with the 
corresponding region of \cite{aleph} obtained for the same set of mSUGRA
parameters. We have used ISAJET version 7.64 \cite{ISAJET} 
for computing the renormalization group evolution of the mSUGRA parameters
and the resulting sparticle spectrum. Since $m_h$ is the most important
parameter in constraining the $m_0 - \MHF$ parameter space,
we  have also used the FeynHiggs programm \cite{FHiggs} to compute the
Higgs masses. In the following
we shall show  that the resulting discrepancy  is not very
serious by comparing the computed  $m_h$ values from the two programs. 

We now impose the UFB3 constraint (inequality \ref{ufb3condition})
and obtain the optimistic upper limit (OL, defined above) 
on $\MHF$ (denoted by $\MHF^{\rm max}$) for each $m_0$. This defines
the lower boundary of region III,  leaving region II
as the only APS. It should  be emphasized that the combination 
of the UFB3 and LEP constraints strengthens the lower bound on
$m_0$ ($m_0^{\rm min}$) as well.  In the absence of the former
constraint  this bound  is $m_0 \gapp 50$ GeV 
(the point of intersection of the upper edge of region I and 
the lower edge of region IV), which
is strengthened to $m_0 \gapp 140$ GeV ( the point of intersection 
of the upper edges of Region
I and II) when the constraints  are combined. The two sets
of constraints, therefore complement each other very well. As the future 
experimental lower bound on $\MHF$ get stronger either from an improved
bound on $m_h$ or from the direct searches  for  gluinos, 
charginos and  neutralinos,
$m_0^{\rm min}$ will also become stronger. As we shall see below this 
corresponds to  indirect lower limits on the masses of the scalar superpartners
 belonging to the first two generations along with some useful  information 
about the other superparticle masses.

The APS for other choices of $\tan\beta$ and sign of $\mu$ are presented
in Fig. 3 and 4. In each case we see that the effectiveness of the 
theoretical and experimental constraints acting 
in tandem is much better than any one of them operating
singly. The lower limit on $m_0$ ($m_0^{\rm min}$) in each case is 
significantly stronger than that obtained from the data alone. 
 


We present in Fig. 5, which is a blown up version of the APS in Fig. 2, 
the uncertainty in $\MHF^{\rm max}$ due to   
the choice of scale for  
$\tan\beta=15, A_0=0, \mu >0$. The thick lines represent  information        
already present in Fig. 2. The optimistic limits (OLs)
as defined above are represented by the thick dash-dotted line. 
The conservative limits (CLs) are given by the line `a'. 
Due to the upward shift of the limits  $m_0^{\rm min}$s  
also get relaxed. 
It is to be noted that the differences between the OLs and the CLs 
become more prominent at larger value of $m_0$.

In Table 1  $m_0^{\rm min}$ and the corresponding $\MHF$
(i.e., the coordinates of the point of intersection of the
upper edges of region I and II which usually determines the
lower bound on the 
slepton mass due to the UFB3 condition in the mSUGRA scenario)
are presented for $A_0=0$ and several choices of $\mu$ and $\tan\beta$.
 The $m_0^{\rm min}$ and $\MHF$ presented in the first two 
columns are obtained from the  light Higgs mass ($m_h$) bound and the
neutralino LSP condition.
The values in the next two columns represent the optimistic constraints 
when  the UFB3 condition is added. The last two columns show
the conservative limits.

Table 2 reflects the uncertainty in the constraints due to different 
methods of computing $m_h$. Here $m_h$ is computed using the two loop corrected
 formulae in the FeynHiggs program \cite{FHiggs}, which are presumably more 
reallistic. The new calculation of the Higgs mass shifts the upper edge
of region I modestly upwards. As a result the relaxation in $m_0^{\rm min}$ 
due to the scale uncertainty is compensated to some extent.
The resulting changes in $m_0^{\min}$ and
 $\MHF$ can be seen by comparing with the correspoding colums in
Table 1.  We see that the changes are $\lapp 10\%$ in all cases.

Using the numbers in Table 2, we predict  in Table 3
the lower limits on the 
slepton masses belonging to  the first two generations. For  low
values of tan $\beta$, even the weaker conservative 
bounds derived using the UFB3 condition 
are significanly larger the corresponding results  
derived from $m_h$ and neutralino LSP constraints only. Detection
of sleptons with masses below the above ranges 
in future experiments would directly signal
the existence of minima deeper than the EW symmetry breaking vacuum.
The chargino masses corresponding the conservative estimate of
the lowest slepton mass are also presented in Table 3.  
Thus even if sleptons with masses  in the ranges shown in Table 3 
are found in future experiments,
but the lighter chargino mass turns out  to be appreciably  larger
, it would also indicate a violation of the UFB3 bound.

Similar lower limits can also be obtained for the squarks 
belonging  to the first two generations. In mSUGRA the squark mass
squared is  given by( apart from the relatively unimportant D-term 
contributions) $m_0^2 + c \MHF^2$, where the dimensionless coefficient
c lies between 6.0 and 6.5 \cite{deboer} for different types of squarks. 
It is clear that     
the main contributions to such limits for relatively low $m_0$ values
 come from the experimental lower limits on $\MHF$.  
The  stroger $m_0^{\rm min}$ as determined from the UFB3 condition, will
in principle  predict a lower  bound stronger than the experimental bound
. But it may be difficult to disitnguish between the two. The shift
in mass limits due to the UFB condition, however, may play an important
role once  precision measurements of these masses are  available.  

We next discuss the predictions of the UFB3 constraint for slepton
masses higher than the lowest allowed values presented in Table 3.
For a fixed slepton mass 
both upper and lower bounds on several sparticle masses  can 
be computed from the allowed ranges of $\MHF$ presented in Figs.  2 - 4.
As an illustration we present in Fig. 5 the contour of left slepton mass
of 300 GeV and 400 GeV. From the points of intersection of contours `e' and 
`f' with the lines representing
the UFB3 bound and the experimental bound on $\MHF$,  we predict 
in Table 4 the conservative and optimistic upper bounds and lower bounds 
on several sparticle masses. It is gratifying to note that the 
uncertainties in the upper limits due to the choice of scale is always
$\lapp 10\%$. If  gluinos happen to be the first sparticles to be 
discovered with a mass 700 GeV, say,  it would predict that
the lower bound on the slepton mass to be 265 GeV, as can be seen from
line `c' of Fig 5.

The upper bounds in Table  4 indicate that the sparticle spectrum 
predicted by the mSUGRA model along with the reqirement of the stability of
the scalar potential can indeed be tested in future experiments.
However, it is desirable that the theoretical predictions, the 
uncertainties due to the choice of renormalization scale in particular,
be further sharpened. In this paper we have followed refs. \cite{gamberini}
and \cite{casas} and assumed that the approximate scale, at which correct
conclusions can be drawn from the tree level scalar potential, is equal 
to $M(\phi)^{\rm max}$ which is the dominant eigen value of the field dependent 
mass matrices. Our conservative limits on SUSY parameters are then obtained if these
limits happen to be stable with respect to the variation of the scale 
within one order of magnitude of the above value.

In practice, however, the above eigen values may have a hierarchy. 
A more rigorous method \cite{gioutsos} which takes into account this
hiearchy and chooses the scale in a step by step fashion using the
decoupling therem \cite{symanzik} is called for. 
A refined estimation of the scale will be particularly important if
the measured masses  are found to be in conflict with the 
theoretical upper bounds after the discovery of SUSY.

Interesting predictions  about some decay properties of sparticles  also
follow from Fig 5. For example,  below the curve marked `b' the 
lighter chargino is the next lightest superparticle (NLSP). Thus a chargino
of mass $\lapp$ 160 GeV is predicted to be the NLSP, since its contour marked
'd' in Fig. 5 
lies below `b' in the entire UFB3 allowed region. Consequently
charginos with  mass in this range
can decay only via  modes
consisting of the LSP and SM particles. All decay channels involving 
on shell superparticles (\eg $\tilde \chi^\pm \to e\tilde \nu$ or 
$\tilde l \nu$)  are ruled out.
In the mSUGRA scenario  the mass of the second lightest neutralino
$(\tilde\chi^0_2)$ closely follows that of the lighter chargino.
Hence similar conclusion about the decay characteristic of
$\tilde\chi^0_2$ can also be drawn.
%
\begin{table}[htbp]
\caption{\sl{ The lowest $m_0$ and the corresponding  $\MHF$ using the spectrum 
generated by the ISASUSY program (see text for further datails). }}
\begin{center}
\begin{tabular}{|c|cc|cc|cc|}
\hline
Choice of parameters   &   ${m_0}_{\rm min}$   &   $\MHF$   &
${m_0}_{\rm min}^{UFB}$     &   $\MHF$              & ${m_0}_{\rm min}^{UFB}$& $\MHF$\\
& (GeV) & (GeV) & (GeV) & (GeV) & (GeV) & (GeV) \\
\hline
$\tan\beta=15,{\rm sign}(\mu)>0$ &45 &200 & 130&198&115&198\\
$\tan\beta=30,{\rm sign}(\mu)<0$ &98 &208 & 148&208&132&208\\
$\tan\beta=44,{\rm sign}(\mu)<0$ &142 &223 & 166&218&150&224\\
\hline
\end{tabular}
\end{center}
\end{table}

\begin{table}[htbp]
\caption{\sl{ The lowest $m_0$ and corresponding  $\MHF$, when 
$m_h$ is computed by the FeynHiggs program. }}
\begin{center}
\begin{tabular}{|c|cc|cc|cc|}
\hline
Choice  of parameters & ${m_0}_{\rm min}$& $\MHF$ &
${m_0}_{\rm min}^{UFB}$& $\MHF$&
${m_0}_{\rm min}^{UFB}$& $\MHF$\\
& (GeV) & (GeV) & (GeV) & (GeV) & GeV & GeV \\
\hline
$\tan\beta=15,{\rm sign}(\mu)>0$ &51 &218  & 140   &  218   &  124  &  218  \\
$\tan\beta=30,{\rm sign}(\mu)<0$  &102  &228 & 159   &  227   &  140  &  227  \\
$\tan\beta=44,{\rm sign}(\mu)<0$  &142 &223 & 170   &  218   &  148  &  222  \\
\hline
\end{tabular}
\end{center}
\end{table}
\begin{table}[htbp]
\caption{\sl{ The  lower limits on slepton masses belonging to the first two 
families and the corresponding chargino masses 
computed using Table 2. The ranges in the mass limits arise 
due to the scale uncertainty.}}
\begin{center}
\begin{tabular}{|c|cc|cc|c|}
\hline
Combination of parameters  &  ~~~~~~~~~~~  ${\msll}^{\rm min}$   &        & ~~~~~~~~~~~~~${\mslr}^{\rm min}$   & & $\MCH$      \\           
                           &  ~~~~~~~~~~~  (GeV)                   &        &   ~~~~~~~~~~~~  (GeV)                &    &(GeV)   \\
\hline
                           & Experiment                            & UFB    &      Experiment                      & UFB &  \\
\hline
$\tan\beta=15,{\rm sign}(\mu)>0$ & 163  & 198 - 208   &  105  &  154 - 167 & 156 \\
$\tan\beta=30,{\rm sign}(\mu)<0$ & 190  & 212 - 225   &  139  &  168 - 184 & 161 \\
$\tan\beta=44,{\rm sign}(\mu)<0$ & 212  & 216 - 229   &  169  &  174 - 192 & 157 \\
\hline
\end{tabular}
\end{center}
\end{table}
\begin{table}[htbp]
\caption{\sl{ The predicted upper and lower limits on sparticle masses 
corresponding to left  slepton
mass $300$($400$) GeV from curve 'e'('f') of Fig. \ref{contour}. The second
and the third
columns contain the input values of $m_0$ and $\MHF$. }}
\begin{center}
\begin{tabular}{|c|c|c|c|c|c|c|c|}
\hline
Limit&$m_0$ & $\MHF$ & $\MCH$ & $\MGL$ & $m_{\tilde \chi^0_1}$ & $m_{\tilde u_L}$   &$ m_{\tilde u_R}$\\
& (GeV) & (GeV) & (GeV) & (GeV) & (GeV) & (GeV) & (GeV) \\
\hline
Lower & 265(375) & 200(200) & 135(136) & 506(514) & 76(76) & 497(557) & 487(549) \\
Upper OL & 195(250) & 334(465) & 245(351) & 793(1074) & 133(189)& 713(963) & 690(931) \\
Upper CL & 175(224) & 355(494) & 262(374) & 838(1135) & 142(201)& 746(1009) & 723(974)\\
\hline
\end{tabular}
\end{center}
\end{table}
At low $\tan\beta$, constraints derived from the UFB3 condition are  not very 
sensative to the sign of $\mu$. In  large $\tan\beta$ scenarios, 
$\MHF^{\rm max}$ is more stringent for a given $m_0$. For $\mu < 0 $, however, 
the constraint becomes moderately 
weaker compared to the  $\mu > 0$ case. 
In the former case the SUSY radiative corrections to the bottom quark
Yukawa coupling makes it smaller compared to 
its magnitude in the latter case. The
effect of this coupling in the renormalization group evolution, which
is quite significant in the large tan $\beta$ scenario, is relatively
subdued for $\mu <$0.
Thus the results presented in Figs 3, 4 and in  subsequent tables 
for $\mu <$ 0 are conservative. We have checked that
   $\MHF^{\rm max}$  may  differ at most by 20 GeV due to the sign of 
$\mu$.

In order to see whether the stability of the potential yields an
improved absolute lower bound on the slepton masses compared to the 
experimental result, a more through scan of the parameter space is needed.
This is beyond the scope of the present paper. However, following comments 
can be made.  

Since the main objective of this letter is to combine the limits of
\cite{aleph} with the theoretical bounds obtained from the UFB3 condition,
we restricted ourselves, as in \cite{aleph} to $A_0 = 0.$
For $A_0 < 0,$ UFB3 bounds are more restrictive since $\MHU$ becomes more 
negative\cite{anirban}. Thus  stronger results for  $\MHF^{\rm max}$ and 
$m_0^{\rm min}$ are expected. For $A_0 >0$ results similar to the $A_0 = 0$
 case are expected, at least for small $|A_0|$. Note that $A_0$ cannot have 
arbitrarily large positive values ($|A_0|<$ 3 $m_0$
from the CCB condition inequality 5 of \cite{casas}). In any case, a correlation between 
the limits on $m_0$, $\MHF$ and positive $A_0$ is expected, 
which may have important
predictions for the third generation of scalar superpartners.
 
Indirect constraints on the $m_0$ - $\MHF$ parameter space have been obtained 
\cite{ellis,cer} from the requirements that 
i) the prediction of the mSUGRA model be consistent
with  the dark matter density of the universe as given by the latest
WMAP data \cite{WMAP}, 
ii) it removes the alleged discrepancy between the measured value of the (g -2)
of the muon \cite{muon} and the standard model prediction or iii) the  
prediction for  the branching ratio of the inclusive process $b \rightarrow 
s \gamma$ be consistent with the measured value
\cite{cleo}. Cerdeno \etal \cite{cer} have also considered the impact of 
the UFB3
constraint on the parameter space. Our conclusions qualitatively agree with 
theirs.

For definiteness we compare
the constaints obtained by Ellis et al \cite{ellis} based on  the latest
WMAP data \cite{WMAP} with our constraints. A precise comparison is difficult
since the constraints in \cite{ellis} are given for different sets of SUSY
parameters. Yet the following qualitative remarks can be made.

From the relevant constrained parameter space
for tan $\beta$ = 10, $A_0$ = 0 and both signs 
of $\mu$ can be found in Fig. (1a) and (1b) of \cite{ellis}. 
Since our constraints are practically insensitive to the sign of $\mu$ for
relatively small tan $\beta$, the constrained parameter space of Fig. 2 of
this letter can be qualitatively 
compared with the results of \cite{ellis}. It is
 interesting to note that for  tan $\beta$ = 10 the dark matter and the
$m_h$ constraints allow only a very narrow region for low $m_0$ 
($\sim$ 100 GeV) 
and relatively large $\MHF$ ($\gapp$ 300 GeV). This is very likely to be in 
conflict with  the UFB3 constraint. Similarly for tan $\beta$ = 35, 
$\mu <$ 0 and $A_0$ = 0, there is a tiny region allowed by both dark matter 
matter and $b \rightarrow s \gamma$ constraints  corresponding 
to $m_0$ between 250 - 350 GeV and $\MHF >$ 700 GeV. For tan $\beta$ = 35
the UFB constraints are expected to be stronger than those presented in Fig. 3.
Qualitatively one can estimate that the above regions will be disfavoured
by the UFB3 constraints. We cannot comment on the compatibility of the 
large $m_0$ regions allowed by indirect constraints with the UFB3 condition 
without a fresh  computation. The incompatibility between  
the post WMAP dark matter constraints and the UFB3 bound
for relatively low tan $\beta$  
was also noted in \cite{cer} and we agree with them. 
On the basis of this observation tan $\beta$ $\lapp$ 20 was ruled out
by the above authors. We note that such conclusions can be evaded by 
introducing a tiny R-parity violation which leaves the LSP essestially stable
for collider experiments but unstable cosmologically. Such small effects can 
be naturally induced by higher dimension operators suppressed by a heavy mass
 scale.  Thus values of
tan $\beta$ less than 20 are still very much relevant for the analysis of 
collider data. 

Qualitatively one can also conclude that significant fractions of the UFB3
allowed parameter space, especially for relatively low $m_0$, will be 
disfavoured by the  $b \rightarrow s \gamma$ constraint for all tan $\beta$
and $\mu < 0$. For $\mu > 0$, on the other hand, the constraint from
 $b \rightarrow s \gamma$ is rather weak. In this scenario 
the region favoured by the 
g-2 constraint has significant overlap with the UFB3 allowed parameter space
for all tan $\beta$.
  
{\bf Acknowlwdgements :} A part of this work was done when AD was visiting
the university of Dortmund under the follow-up visit programme of the
Alexander von Humboldt Foundation. He thanks Professor E. A. Paschos for
 hospitatlity. 
\begin{figure}[htb]
\centerline{
\psfig{file=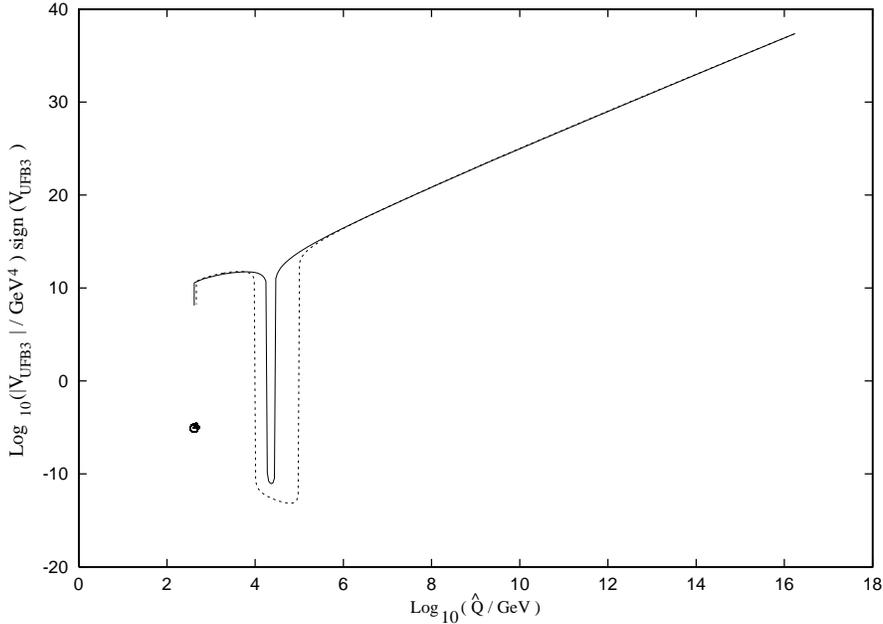,width=12cm,angle=270}}
\caption{\sl{The variation of  $Log_{10}(|V_{UFB3}|) \times$
             sign ($V_{UFB3}$) 
              with the logarithm of the scale $\QA$ for 
             $\MHF = 220$ GeV (solid line) and  250 GeV (dotted line).
             The points denoted by the solid triangle and the open circle 
             indicate the  value of  $Log_{10}(|V_0^{\rm min}|) 
             \times$ sign ($V_0^{\rm min}$)  
              for $\MHF$ = 220 and 250 GeV 
             respectively. 
            We set $m_0=140$ GeV, $A_0=0$, $\tan\beta= 15$ and $\mu>0$.}}
\label{mhf1}
\end{figure}
\begin{figure}[htb]
\centerline{
\psfig{file=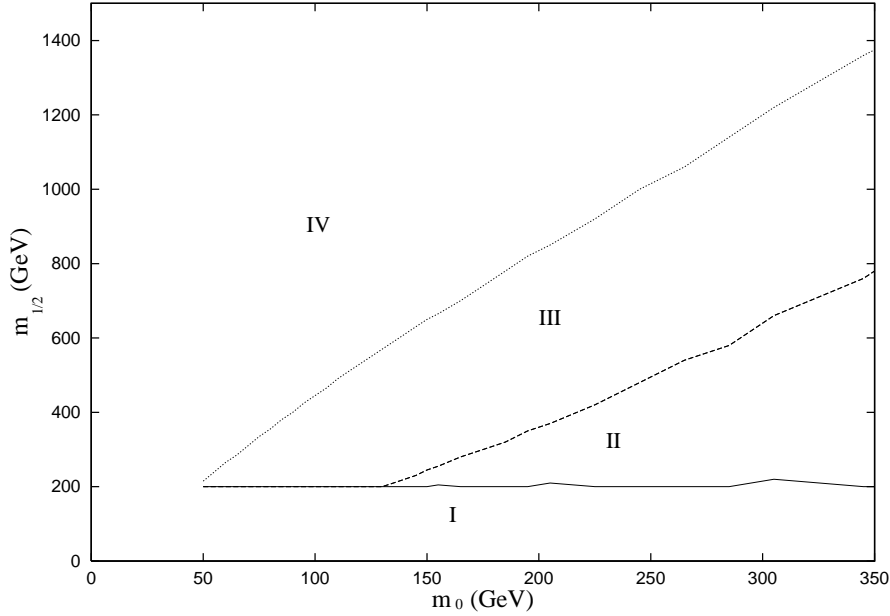,width=12cm,angle=270}}
\caption{\sl{The allowed parameter space for
            $A_0=0$, $\tan\beta= 15$ and $\mu>0$.
            The region II is allowed. The region III is ruled out from UFB3 constraint.
The region I and IV are ruled out from experiment and neutralino LSP condition.}}
    \label{tan15a00mu1}
\end{figure}

\begin{figure}[htb]
\centerline{
\psfig{file=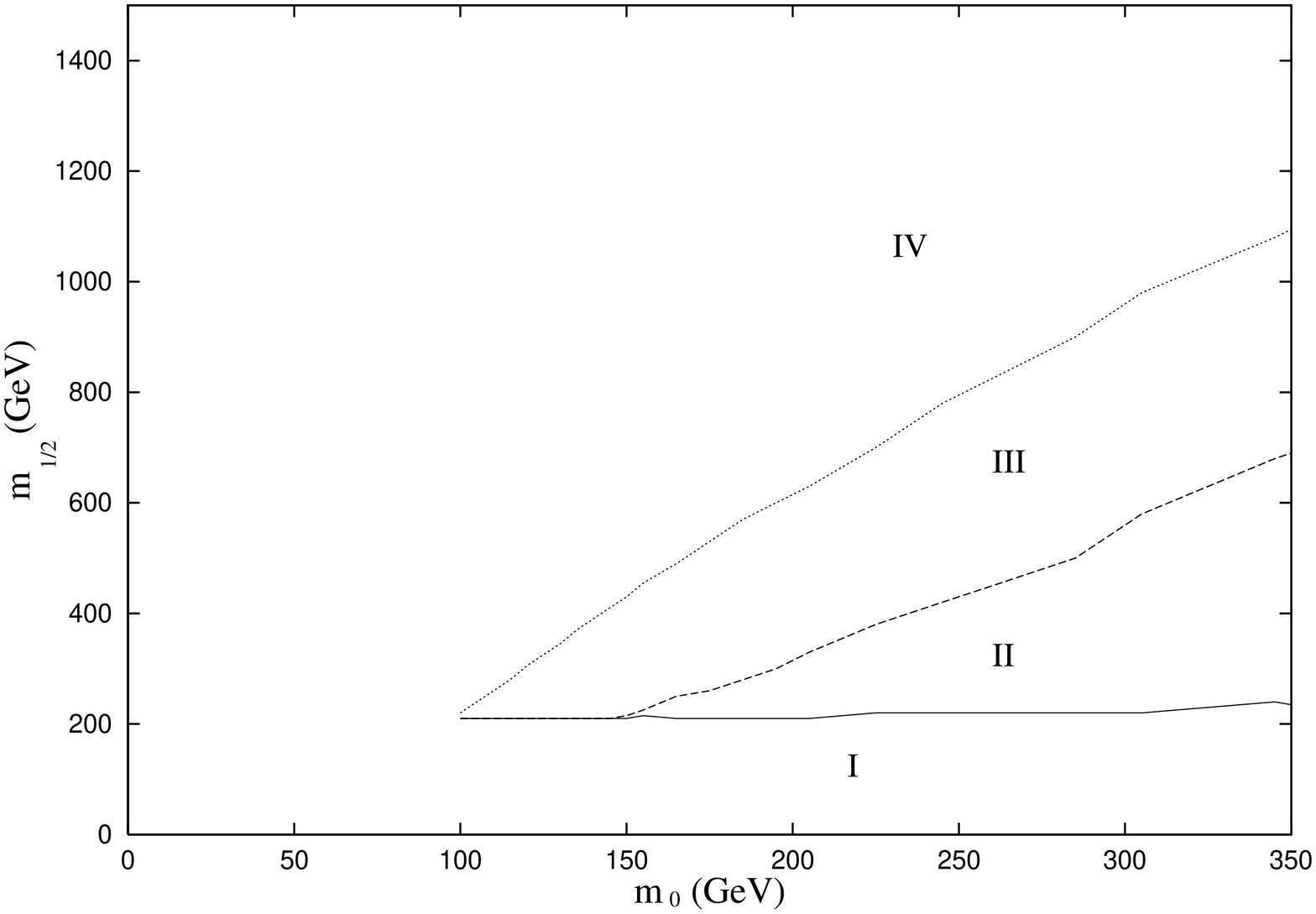,width=12cm,angle=270}}
\caption{\sl{The same as Fig. 2 with
            $A_0=0$, $\tan\beta= 30$ and $\mu<0$.
            }}
    \label{tan30a00mu_1}
\end{figure}

\begin{figure}[htb]
\centerline{
\psfig{file=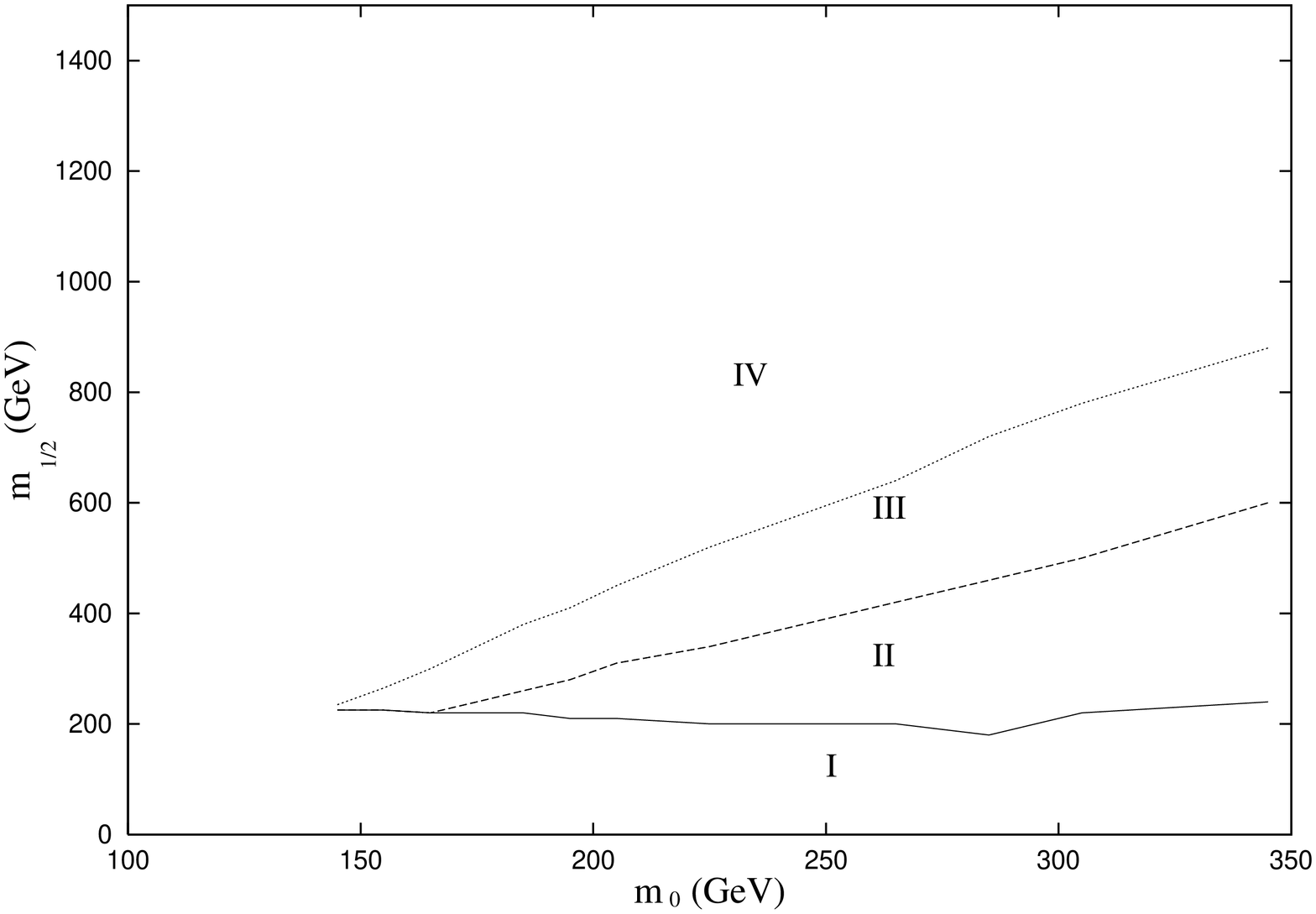,width=12cm,angle=270}}
\caption{\sl{The same as Fig. 2 with
            $A_0=0$, $\tan\beta= 44$ and $\mu<0$.
            }}
    \label{tan44a00mu_1}
\end{figure}
\begin{figure}[htb]
\centerline{
\psfig{file=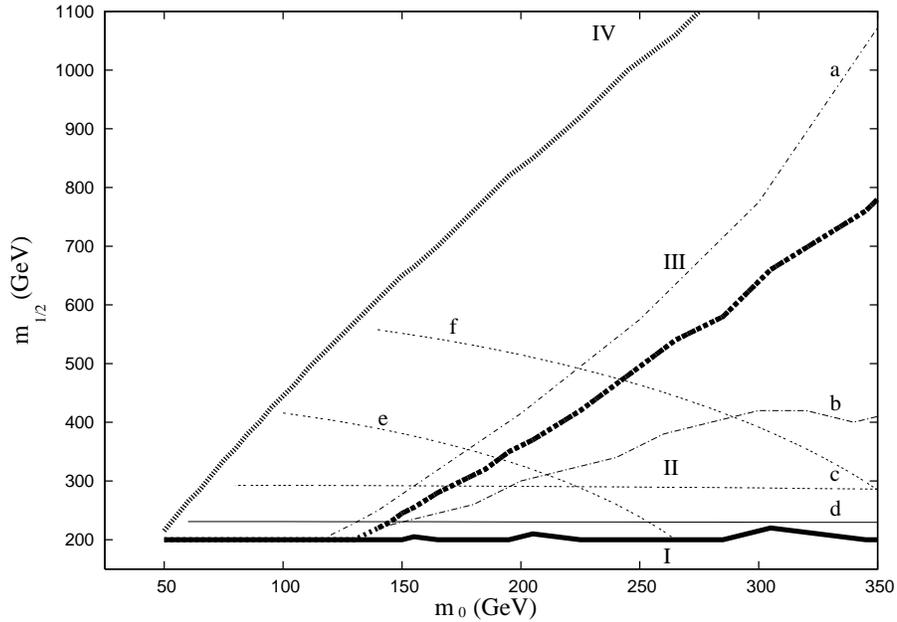,width=12cm,angle=270}}
\caption{\sl{
The blown up version of the 
allowed parameter space  in Fig. \ref{tan15a00mu1}.
The curve labelled by 'a' represents the UFB3 bound after
including the scale uncertainty, in
the region under the curve  'b' the lighter  chargino is the NLSP.
The curves 'c' and 'd' are the contours for $\MGL=$700 GeV
and $\MCH=$160 GeV; the curves 'e' and 'f' are  
fixed slepton mass contours for  300 and 400 GeV respectively.
            }}
    \label{contour}
\end{figure}



\end{document}